\documentclass[twocolumn,english,prl]{revtex4}
\usepackage[T1]{fontenc}
\usepackage[latin9]{inputenc}
\usepackage{booktabs}
\usepackage{bm}
\usepackage{amsmath}
\usepackage{graphicx}

\makeatletter

\providecommand{\tabularnewline}{\\}

\usepackage{bm}
\usepackage{times}

\makeatother

\usepackage{babel}

\begin{document}

\title{Effective Interacting Hamiltonian and Pairing Symmetry of LaOFeAs}

\author{Junren Shi}

\email{jrshi@aphy.iphy.ac.cn}

\affiliation{Institute of Physics and ICQS, Chinese Academy of Sciences, Beijing
100190, China}

\begin{abstract}
We establish the general form of effective interacting Hamiltonian
for LaOFeAs system based on the symmetry consideration. The peculiar
symmetry property of the electron states yields unusual form of electron-electron
interaction. Based on the general effective Hamiltonian, we determine
all the ten possible pairing states. More physical considerations
would further reduce the list of the candidates for the pairing state.
\end{abstract}
\maketitle
The recent discovery of the new family of iron based high-$T_{c}$
superconductors~\citep{Kamihara2008,Chen2008,Wen2008,Ren2008a,Takahashi2008}
has attracted intensive experimental~\citep{Cruz2008,Dong2008} and
theoretical interests. Although at the current stage little is known
for its microscopic origin, theory has made great advances in understanding
the electronic structures~\citep{Singh2008,Zhang2008a,Kuroki2008,Haule2008}.
In particular, a number of pairing mechanisms and pairing symmetries
have been proposed~\citep{Singh2008,Dai2008,Kuroki2008,Han2008,Qi2008,Li2008,Lee2008,Weng2008,Seo2008}.
In most of these studies, the microscopic Hamiltonian adopted is deterministic
for the outcome of theory. It is thus desirable to know the general
form of the interacting Hamiltonian allowed by the symmetry of the
system, upon which the possible pairing states can be systematically
analyzed.

In this paper, we establish the effective interacting Hamiltonian
for LaOFeAs system based on the general symmetry consideration. The
peculiar symmetry property of electron states near $M$-point yields
unusual form of electron-electron ($e$-$e$) interaction. Based on
the general effective Hamiltonian, we determine all the possible pairing
states allowed by the symmetry. The stability of these pairing states
against the band energy splitting and the on-site Coulomb repulsion
is discussed. The analysis is general enough to be useful for other
systems with the similar electronic structure.

\begin{figure}
\includegraphics[width=1\columnwidth]{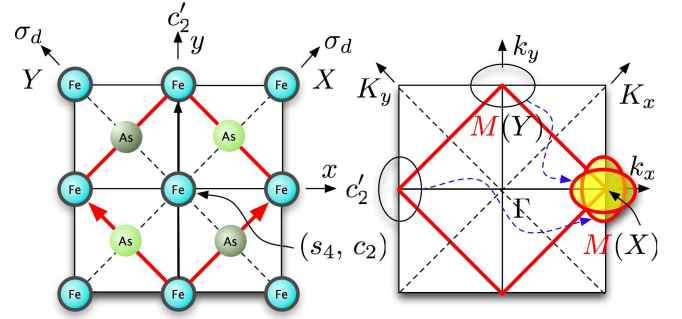}

\caption{\label{fig:Structure} Left: structure of FeAs plane. The dark and
light As atoms reside below and above the plane of Fe-square lattice,
respectively. The red (thick) square indicates the $\sqrt{2}\times\sqrt{2}$
primitive cell containing two Fe atoms. Note that it is also possible
to define a $1\times1$ primitive cell containing only one Fe atom,
using a generalized Bloch theorem~\citep{Lee2008}. The symmetry
axes for $D_{2d}$ symmetry operations are also indicated. Right:
The real Brillouin zone (red thick square) and the extended Brillouin
zone corresponding to the $1\times1$ primitive cell (the larger black
square). The latter can be folded into the real Brillouin zone, yielding
the two elliptic Fermi pockets at $M$-point (shaded) originally located
at $X$ and $Y$-points.}

\end{figure}

\textit{Structure and symmetry.} The structure of LaOFeAs consists
of the alternating layers of FeAs and LaO planes. The first principles
calculations reveal the dominant role of the two-dimensional FeAs
planes in electron conduction~\citep{Singh2008,Zhang2008a,Kuroki2008}.
Fig.~\ref{fig:Structure} shows the schematic structure of the FeAs
plane. The full symmetry group for the system is of $P4/nmm$~\citep{Wang2008,Wan2008}.
For our purpose, it is sufficient to consider its symmorphic subgroup
$P\bar{4}2m$, which is a semi-direct-product of the point group $D_{2d}$
and the lattice translational group. Figure~\ref{fig:Structure}
shows the symmetry axes of the eight symmetry operations of $D_{2d}$.
The point group has four one-dimensional irreducible representations:
$A_{1}$ ($x^{2}+y^{2}$), $A_{2}$ ($xy(x^{2}-y^{2})$), $B_{1}$
($x^{2}-y^{2}$), $B_{2}$ ($xy$), and a two-dimensional representation
$E$ ($(xz,yz)$). 

The most notable feature of LaOFeAs system is its multi-orbit nature.
The Bloch bands form five small Fermi pockets: three hole pockets
at $\Gamma$-point, and two electron pockets at $M$-point. It was
suggested that the hole pockets shrink and disappear upon doping,
leaving the two electron pockets responsible for the superconductivity~\citep{Singh2008}.
The two electron pockets are of Fe-$d$ orbit origin, and the Bloch
states at the $M$-point are essentially $d_{xz}$ and $d_{yz}$.
Departing from the $M$-point, these states strongly hybridize with
$d_{xy}$ orbit, giving rise to the two elliptically shaped Fermi
pockets~%
\footnote{Due to the strong hybridization, a minimal model to describe the band
structure near $M$-point should include at least $d_{xz}$, $d_{yz}$
and $d_{xy}$ orbits~\citep{Lee2008}.%
}. 

The electron states near $M$-point have peculiar symmetry property.
The two degenerated Bloch states at $M$-point span the subspace for
the two-dimensional irreducible representation $E$ of $D_{2d}$.
Departing from the $M$-point, the Bloch states $\varphi_{i}(\bm{k},\bm{r})$
($i=1,2$) of the two bands will be transformed by:

\begin{equation}
\hat{\alpha}\varphi_{i}(\bm{k},\bm{r})=\varphi_{i}(\bm{k},\hat{\alpha}^{-1}\bm{r})=\sum_{j=1,2}D_{ij}^{E}(\alpha)\varphi_{j}(\hat{\alpha}\bm{k},\bm{r}),\label{eq:transformation}\end{equation}
where $\hat{\alpha}\in D_{2d}$ is a symmetry operation, and $D_{ij}^{E}(\alpha)$
is the two-dimensional irreducible representation matrix for $(xz,yz)$.
The transformation property can be established by generalizing the
usual symmetry argument for Bloch wave-functions with the proper assignment
of the band index to the two states at each $\bm{k}$ point, so that
the resulting wave functions are continuous functions of $\bm{k}$.
Note that Eq.~(\ref{eq:transformation}) is possible because of the
presence of a generalized translational symmetry: the system is invariant
under the transformations $T_{x}P_{z}$ and $T_{y}P_{z}$, where $T_{x}$
($T_{y}$) is the translation along $x$ ($y$) direction by the Fe-Fe
distance and $P_{z}$ is the reflection $z\rightarrow-z$~\citep{Lee2008}.
With the symmetry, one can define a reduced $1\times1$ primitive
cell containing only one Fe atom, and the corresponding extended Brillouin
zone, as shown in Fig.~\ref{fig:Structure}. The two overlapping
Fermi pockets at $M$-point in the real Brillouin zone actually originate
from the Fermi pockets located at the two non-equivalent points $X$
and $Y$ of the extended Brillouin zone. Without such a symmetry,
the hybridization between different $d$-orbits would re-organize
them into two separated Fermi pockets that transform within themselves,
instead of the two-dimensional representation shown in Eq.~(\ref{eq:transformation}).
Note that the presence of the generalized translational symmetry does
not prohibit interband coupling through $e$-$e$ interaction. 

\textit{Effective electron-electron interaction.} Equation (\ref{eq:transformation})
is one of the peculiarities of LaOFeAs system. It is interesting to
see how the unique electron structure has effect on the $e$-$e$
interaction. We thus focus on the interaction amongst the two electron
pockets, which can be written as:\begin{multline}
H_{\mathrm{I}}=\sum_{is}V_{i_{1}i_{2};i_{3}i_{4}}^{s_{1}s_{2};s_{3}s_{4}}(\bm{k},\bm{k}^{\prime})\\
\times a_{i_{1}s_{1}}^{\dagger}(\bm{k})a_{i_{2}s_{2}}^{\dagger}(-\bm{k})a_{i_{3}s_{3}}(-\bm{k}^{\prime})a_{i_{4}s_{4}}(\bm{k}^{\prime}),\label{eq:HI}\end{multline}
where $a_{is}^{\dagger}(\bm{k})$ and $a_{is}(\bm{k})$ are the creation
and annihilation operators for band $i=1,2$ and spin index $s=\uparrow\downarrow$.
We only consider $e$-$e$ interaction relevant to the superconductivity
between a pair of electrons with the zero total momentum. The $e$-$e$
interaction should be interpreted as an effective one, including contributions
not only from the direct Coulomb interaction, but also from the effective
interaction mediated by other degrees of freedom such as phonon, or
the renormalization effect due to the projection of the high-energy
sectors. 

Next we explicitly construct the general interacting Hamiltonian invariant
under the symmetry operations. From Eq.~(\ref{eq:transformation}),
it is easy to see that the annihilation operator $a_{i}(\bm{k})$
transforms by $\hat{\alpha}a_{i}(\bm{k})\hat{\alpha}^{-1}=\sum_{j}a_{j}(\hat{\alpha}^{-1}\bm{k})D_{ji}^{E}(\alpha)$,
here we ignore the spin indexes for the moment. As a result, the transformations
of $a_{i_{1}}^{\dagger}(\bm{k})a_{i_{2}}^{\dagger}(-\bm{k})a_{i_{3}}(-\bm{k}^{\prime})a_{i_{4}}(\bm{k}^{\prime})$
under the symmetry operations belong to the direct-product representation
$E\times E\times E\times E$, which can be decomposed to the one-dimensional
irreducible representations $4A_{1}\oplus4A_{2}\oplus4B_{1}\oplus4B_{2}$. 

The decomposition can be explicitly constructed by the following procedures.
First, the two-particle annihilation operator $a_{i}(-\bm{k})a_{j}(\bm{k})$
can be organized to different orbital pairing channels: $\Psi_{0}(\bm{k})\equiv a_{i}(-\bm{k})\tau_{ij}^{0}a_{i}(\bm{k})\sim A_{1}$,
$\Psi_{1}(\bm{k})\equiv a_{i}(-\bm{k})\tau_{ij}^{1}a_{j}(\bm{k})\sim B_{2}$,
$\Psi_{2}(\bm{k})\equiv a_{i}(-\bm{k})\tau_{ij}^{2}a_{j}(\bm{k})\sim A_{2}$,
$\Psi_{3}(\bm{k})\equiv a_{i}(-\bm{k})\tau_{ij}^{3}a_{j}(\bm{k})\sim B_{1}$,
where $\tau^{\alpha}$ ($\alpha=1,2,3)$ are the Pauli-matrices, and
$\tau^{0}$ is the unity matrix~\citep{Wan2008}. The symbol $\sim$
indicates the transformation property of the operators. For instance,
$\Psi_{0}(\bm{k})\sim A_{1}$ means $\hat{\alpha}\Psi_{0}(\bm{k})\hat{\alpha}^{-1}=D^{A_{1}}(\alpha)\Psi_{0}(\hat{\alpha}^{-1}\bm{k})$,
where $D^{A_{1}}(\alpha)$ is the transformation coefficient of the
irreducible representation $A_{1}$. Using these bases, the effective
$e$-$e$ interaction can be written as:\begin{equation}
H_{\mathrm{I}}=\sum_{\bm{k}\bm{k}^{\prime}}\Psi^{\dagger}(\bm{k})\left[\begin{array}{cccc}
v_{00}^{A_{1}} & v_{01}^{B_{2}} & v_{02}^{A_{2}} & v_{03}^{B_{1}}\\
* & v_{11}^{A_{1}} & v_{12}^{B_{1}} & v_{13}^{A_{2}}\\
* & * & v_{22}^{A_{1}} & v_{23}^{B_{2}}\\
* & * & * & v_{33}^{A_{1}}\end{array}\right]\Psi(\bm{k}^{\prime}),\label{eq:reduced HI}\end{equation}
where $\Psi(\bm{k})\equiv[\Psi_{0},\Psi_{1},\Psi_{2},\Psi_{3}]^{T}$,
and all the matrix elements $v_{ij}$ are functions of $(\bm{k},\bm{k}^{\prime})$,
and have symmetry $v_{ij}(\bm{k},\bm{k}^{\prime})=v_{ji}^{*}(\bm{k}^{\prime},\bm{k})$.
To make the Hamiltonian invariant under the point group operations,
$v_{ij}(\bm{k})$ must have certain transformation property. For instance,
$\Psi_{1}^{\dagger}(\bm{k})\Psi_{2}(\bm{k}^{\prime})\sim B_{2}\times A_{2}=B_{1}$,
i.e., $\hat{\alpha}[\Psi_{1}^{\dagger}(\bm{k})\Psi_{2}(\bm{k}^{\prime})]\hat{\alpha}^{-1}=D^{B_{1}}(\alpha)\Psi_{1}^{\dagger}(\hat{\alpha}^{-1}\bm{k})\Psi_{2}(\hat{\alpha}^{-1}\bm{k}^{\prime})$.
To have an invariant partial Hamiltonian $\sum_{\bm{k}\bm{k}^{\prime}}v_{12}(\bm{k},\bm{k}^{\prime})\Psi_{1}^{\dagger}(\bm{k})\Psi_{2}(\bm{k}^{\prime})$,
$v_{12}$ must have $v_{12}(\hat{\alpha}^{-1}\bm{k},\hat{\alpha}^{-1}\bm{k}^{\prime})=D^{B_{1}}(\alpha)v_{12}(\bm{k},\bm{k}^{\prime})$,
i.e., $v_{12}\sim B_{1}$. Following the same procedure, the symmetry
properties of all matrix elements can be determined. They are indicated
by the superscripts of $v_{ij}$ in Eq.~(\ref{eq:reduced HI}).

To incorporate the spin indexes into the effective interaction, we
 generalize the pair annihilation operators $\Psi_{\alpha}(\bm{k})\rightarrow\Psi_{\alpha,a}(\bm{k})\equiv a_{is}(-\bm{k})\tau_{ij}^{\alpha}\tau_{ss^{\prime}}^{a}a_{js^{\prime}}(\bm{k})$,
where $a$ denotes the spin pairing channel. When the spin-orbit coupling
is negligible, the effective $e$-$e$ interaction has the form~\citep{Nakajima1973}:\begin{multline}
H_{\mathrm{I}}=\sum_{\bm{k}\bm{k}^{\prime}}\Psi_{\mbox{},2}^{\dagger}(\bm{k})V^{(s)}(\bm{k},\bm{k}^{\prime})\Psi_{\mbox{},2}(\bm{k}^{\prime})\\
+\sum_{a\ne2}\Psi_{\mbox{},a}^{\dagger}(\bm{k})V^{(t)}(\bm{k},\bm{k}^{\prime})\Psi_{\mbox{},a}(\bm{k}^{\prime})\label{eq:HI-s}\end{multline}
where $\Psi_{,a}(\bm{k})\equiv[\Psi_{0,a},\Psi_{1,a},\Psi_{2,a},\Psi_{3,a}]^{T}$,
and the matrices $V^{(s)}(\bm{k},\bm{k}^{\prime})$ and $V^{(t)}(\bm{k},\bm{k}^{\prime})$
denote the effective $e$-$e$ interactions in the spin-singlet $(s)$
and triplet $(t)$ channels, respectively. Both matrices have the
same structure as the one presented in Eq.~(\ref{eq:reduced HI}).
Pauli exclusion principle imposes the further constraints onto the
matrix elements: $v_{\alpha\beta}^{(s)}(-\bm{k},\bm{k}^{\prime})=g(\alpha)v_{\alpha\beta}^{(s)}(\bm{k},\bm{k}^{\prime})$,
$v_{\alpha\beta}^{(s)}(\bm{k},-\bm{k}^{\prime})=g(\beta)v_{\alpha\beta}^{(s)}(\bm{k},\bm{k}^{\prime})$,
and $v_{\alpha\beta}^{(t)}(-\bm{k},\bm{k}^{\prime})=-g(\alpha)v_{\alpha\beta}^{(t)}(\bm{k},\bm{k}^{\prime})$,
$v_{\alpha\beta}^{(t)}(\bm{k},-\bm{k}^{\prime})=-g(\beta)v_{\alpha\beta}^{(t)}(\bm{k},\bm{k}^{\prime})$,
where $g(\alpha)=-1$ ($1$) for $\alpha=2$ ($\ne2$). The matrix
elements $v_{\alpha\beta}(\bm{k},\bm{k}^{\prime})$ can be further
expanded as: \begin{equation}
v_{\alpha\beta}^{R}(\bm{k},\bm{k}^{\prime})=\sum_{nn^{\prime},R\in R_{1}\times R_{2}}v_{nn^{\prime},\alpha\beta}^{R_{1}\mbox{-}R_{2}}f_{n}^{R_{1}}(\bm{k})f_{n^{\prime}}^{R_{2}}(\bm{k}^{\prime}),\label{eq:fusion}\end{equation}
where $ $$f_{n}^{R}(\bm{k})$ denotes a complete set of functions
that transform by the irreducible representation $R$. The summation
should be run through all the possible combinations of $R_{1}$ and
$R_{2}$ which yield $R\in R_{1}\times R_{2}$. For instance, for
$v_{03}^{B_{1}}$, there are five possible combinations, $A_{1}\times B_{1}$,
$B_{1}\times A_{1}$, $B_{2}\times A_{2}$, $A_{2}\times B_{2}$,
$E\times E$, based on the product rules of the group. 

To be more explicit, we  construct an expression using only the lowest
order polynomials for each irreducible representation: $s\sim1\sim A_{1}$,
$g\sim k_{x}k_{y}(k_{x}^{2}-k_{y}^{2})\sim A_{2}$, $d_{x^{2}-y^{2}}\sim k_{x}^{2}-k_{y}^{2}\sim B_{1}$,
$d_{xy}\sim k_{x}k_{y}\sim B_{2}$, $[p_{x},p_{y}]\sim[k_{x},k_{y}]\sim E$.
Such an expression contains the full information of symmetry. It could
also be a good approximation for LaOFeAs system since all its Fermi
pockets are small, and the higher ($n$-th) order contributions are
scaled by $(k_{F}a)^{n}$, where $a$ is the scattering length of
the effective interaction, and $k_{F}$ is the Fermi wave vector.
The explicit form of $v_{\alpha\beta}^{(s)}(\bm{k},\bm{k}^{\prime})$
reads:\begin{align}
v_{ii}^{(s)}= & v_{i}^{s}+v_{i}^{d_{1}}(k_{x}^{2}-k_{y}^{2})(k_{x}^{\prime2}-k_{y}^{\prime2})+v_{i}^{d_{2}}k_{x}k_{y}k_{x}^{\prime}k_{y}^{\prime}\label{eq:vii}\\
 & +v_{i}^{g}k_{x}k_{y}(k_{x}^{2}-k_{y}^{2})k_{x}^{\prime}k_{y}^{\prime}(k_{x}^{\prime2}-k_{y}^{\prime2}),\,\,\, i=0,\,1,\,3,\nonumber \\
v_{01}^{(s)}= & v_{01}^{d_{2}\mbox{-}s}k_{x}k_{y}+v_{01}^{d_{1}\mbox{-}g}(k_{x}^{2}-k_{y}^{2})k_{x}^{\prime}k_{y}^{\prime}(k_{x}^{\prime2}-k_{y}^{\prime2})+\\
 & v_{01}^{s\mbox{-}d_{2}}k_{x}^{\prime}k_{y}^{\prime}+v_{01}^{g\mbox{-}d_{1}}k_{x}k_{y}(k_{x}^{2}-k_{y}^{2})(k_{x}^{\prime2}-k_{y}^{\prime2}),\nonumber \\
v_{03}^{(s)}= & v_{03}^{s\mbox{-}d_{1}}\left(k_{x}^{\prime2}-k_{y}^{\prime2}\right)+v_{03}^{d_{2}\mbox{-}g}k_{x}k_{y}k_{x}^{\prime}k_{y}^{\prime}(k_{x}^{\prime2}-k_{y}^{\prime2})+\label{eq:v03}\\
 & v_{03}^{d_{1}\mbox{-}s}\left(k_{x}^{2}-k_{y}^{2}\right)+v_{03}^{g\mbox{-}d_{2}}k_{x}k_{y}(k_{x}^{2}-k_{y}^{2})k_{x}^{\prime}k_{y}^{\prime},\nonumber \\
v_{13}^{(s)}= & v_{13}^{s\mbox{-}g}k_{x}^{\prime}k_{y}^{\prime}\left(k_{x}^{\prime2}-k_{y}^{\prime2}\right)+v_{13}^{d_{1}\mbox{-}d_{2}}\left(k_{x}^{2}-k_{y}^{2}\right)k_{x}^{\prime}k_{y}^{\prime}+\label{eq:v13}\\
 & v_{13}^{g\mbox{-}s}k_{x}k_{y}\left(k_{x}^{2}-k_{y}^{2}\right)+v_{13}^{d_{2}\mbox{-}d_{1}}k_{x}k_{y}\left(k_{x}^{\prime2}-k_{y}^{\prime2}\right),\nonumber \\
v_{22}^{(s)}= & v_{2}^{p}\left(k_{x}k_{x}^{\prime}+k_{y}k_{y}^{\prime}\right),\label{eq:v22}\\
v_{23}^{(s)}= & v_{12}^{(s)}=v_{02}^{(s)}=0.\label{eq:v23}\end{align}
Similarly, the explicit expression for $v_{\alpha\beta}^{(t)}$ reads:\begin{align}
v_{22}^{(t)}= & v_{2}^{s}+v_{2}^{d_{1}}(k_{x}^{2}-k_{y}^{2})(k_{x}^{\prime2}-k_{y}^{\prime2})+v_{2}^{d_{2}}k_{x}k_{y}k_{x}^{\prime}k_{y}^{\prime}\nonumber \\
 & +v_{2}^{g}k_{x}k_{y}(k_{x}^{2}-k_{y}^{2})k_{x}^{\prime}k_{y}^{\prime}(k_{x}^{\prime2}-k_{y}^{\prime2}),\label{eq:vt22}\\
v_{ii}^{(t)}= & v_{i}^{p}\left(k_{x}k_{x}^{\prime}+k_{y}k_{y}^{\prime}\right),\,\,\, i=0,1,3,\label{eq:vtii}\\
v_{01}^{(t)}= & v_{01}^{p\mbox{-}p}\left(k_{x}k_{y}^{\prime}+k_{x}^{\prime}k_{y}\right),\label{eq:vt01}\\
v_{03}^{(t)}= & v_{03}^{p\mbox{-}p}\left(k_{x}k_{x}^{\prime}-k_{y}k_{y}^{\prime}\right),\label{eq:vt03}\\
v_{13}^{(t)}= & v_{13}^{p\mbox{-}p}\left(k_{x}k_{y}^{\prime}-k_{y}k_{x}^{\prime}\right),\label{eq:vt13}\\
v_{23}^{(t)}= & v_{12}^{(t)}=v_{02}^{(t)}=0.\label{eq:vt23}\end{align}

\textit{Possible pairing symmetries.} Equations (\ref{eq:HI-s})--(\ref{eq:vt23})
provide the basis for deducing the possible pairing symmetries. Basically,
the interacting Hamiltonian Eq.~(\ref{eq:HI-s}) describes how the
electron pairs in different orbital and spin pairing channels are
coupled, while Eq.~(\ref{eq:fusion}) (or (\ref{eq:vii})--(\ref{eq:vt23}))
dictates the coupling between the different momentum pairing symmetries.
The pairing instability of the system can be determined by considering
a Cooper pair out of the Fermi sea. The Cooper equation reads~%
\footnote{The equation is actually a linearized BCS gap equation. It is the
exact gap equation at the limit of the vanishing gap. %
},\begin{multline}
h_{\alpha\beta}(\bm{k})\psi_{\alpha,a}(\bm{k})+\sum_{\beta,\bm{k}^{\prime}}v_{\alpha\beta}^{(a)}(\bm{k},\bm{k}^{\prime})w_{\beta}(\bm{k}^{\prime})\psi_{\beta,a}(\bm{k}^{\prime})\\
=E\psi_{\alpha,a}(\bm{k})\label{eq:Cooper}\end{multline}
where $h_{\alpha\beta}(\bm{k})$ is the kinetic energy of the Cooper
pair, and has the non-vanishing elements: $h_{00}(\bm{k})=h_{33}(\bm{k})=|\xi_{1}(\bm{k})|+|\xi_{2}(\bm{k})|$,
$h_{11}(\bm{k})=h_{22}(\bm{k})=|\xi_{1}(\bm{k})+\xi_{2}(\bm{k})|$,
$h_{03}(\bm{k})=h_{30}(\bm{k})=|\xi_{1}(\bm{k})|-|\xi_{2}(\bm{k})|$,
where $\xi_{i}(\bm{k})=\epsilon_{i}(\bm{k})-\mu$ is the single electron
band energy relative to the Fermi surface. Note that the interband
Cooper pairs cannot exist in the non-overlapping areas of the two
Fermi pockets, which are excluded from contributing the interband
pairing potential by the measure $w_{\beta}(\bm{k)}$: $w_{0}(\bm{k})=w_{3}(\bm{k})=1$,
$w_{1}(\bm{k})=w_{2}(\bm{k})=\theta(\xi_{1}(\bm{k})\xi_{2}(\bm{k}))$,
where $\theta(x)$ is the Heaviside function. A bounding state solution
of the equation signifies the instability of the normal Fermi liquid,
and the resulting superconducting state will have the order parameter
$\Delta_{ij,ss^{\prime}}(\bm{k})\sim\tau_{ij}^{\alpha}\tau_{ss^{\prime}}^{a}|\xi_{i}(\bm{k})+\xi_{j}(\bm{k})|\psi_{\alpha,a}(\bm{k})$,
approximately.

It is easy to see both $h_{\alpha\beta}$ and $w_{\beta}(\bm{k})$
do not change the symmetry characteristics of Eq.~(\ref{eq:Cooper})
from the one dictated by $v_{\alpha\beta}^{(a)}$. It is thus straightforward
to use Eq.~(\ref{eq:vii})--(\ref{eq:vt23}) to determine the possible
pairing states of the system. It can be readily observed that the
pairing states must have the definite parities in exchanging spin/orbit/momentum
indexes, as the states with the different parities do not mix. A closer
analysis reveals the particular way of mixing between the different
pairing symmetries and the orbital pairing channels. For instance,
in the spin singlet channel, an $s$-wave component in orbital pairing
channel $0$ ($\psi_{0}\sim s$) will induce $\psi_{1}\sim d_{xy}$
(by $v_{01}^{(s)}$) and $\psi_{3}\sim d_{x^{2}-y^{2}}$ (by $v_{03}^{(s)}$).
These states form a closed subspace for an eigenstate of Eq.~(\ref{eq:Cooper}).
The corresponding superconductivity order parameter has the form~\citep{Seo2008}:\begin{equation}
\Delta_{ij}(\bm{k})\sim\sum_{\alpha}\tau_{ij}^{\alpha}\psi_{\alpha}=\left[\begin{array}{cc}
s+d_{x^{2}-y^{2}} & d_{xy}\\
d_{xy} & s-d_{x^{2}-y^{2}}\end{array}\right].\end{equation}
The similar analyses can be carried out for all other possible combinations.
A complete list of all possible pairing states is presented in Table~\ref{tab:pairing}.

More physical considerations may further reduce the list of candidates.
One of the most important factors is the band energy splitting $\xi_{1}(\bm{k})-\xi_{2}(\bm{k})$,
which acts like a {}``Zeeman'' field and will suppress the interband
pairings. Its adverse effects to the different interband pairing states
have the relative strengths $d_{x^{2}-y^{2}}>s\approx p\approx g>d_{xy}$.
The strong band splitting tends to suppress the pure interband pairing
states (5) and (7)--(10), although the state (9) may be more robust
than the others in the group. On the other hand, those states mixing
the interband and intraband pairings, i.e., the states (1)--(4) and
(6), may not be as sensitive, as long as the pairing is dominated
by the intraband attractive interaction. Another potential factor
is the strong on-site Coulomb repulsion, which would suppress $s$-wave
components in states (1)--(3) and (7). As a result, the states (1)--(3)
will be dominated by the $d$-wave pairing symmetries. Finally, the
smallness of the Fermi pockets in LaOFeAs system may also have implications
on the possible pairing forms: the screening of the bare $e$-$e$
interaction by the Fermi gas of small $k_{F}$ will render the effective
interaction spatially extended. It is thus unlikely for LaOFeAs system
to form spatially localized bond-like-pairings, as that happens in
high-$T_{c}$ cuprates~\citep{Wan2008}. Moreover, in the limit of
weak coupling ($k_{F}a\ll1$), one may expect that the attractive
$e$-$e$ interaction in the $p$-wave channel, if presents, would
be stronger than that in the $d$-wave channels. Note that in many
microscopic models, the $p$-wave channel is absent, due to the improper
adoption of a two-orbit model for describing the band structure near
the $M$-point~\citep{Lee2008}, while the symmetry argument clearly
suggests its presence. The possibility of $p$-wave pairing state
(i.e., the state (6)) has been demonstrated in Ref.~\citep{Lee2008},
using a microscopic model including three $d$-orbits and on-site
Hund's coupling.

\begin{table}
\begin{tabular}{ccccc}
\toprule 
 & $\Delta_{ij}$ & $P_{\mathrm{orbit}}$  & $P_{\mathrm{spin}}$ & I.R.\tabularnewline
\midrule
\midrule 
1 & $\left(\begin{array}{cc}
s+d_{x^{2}-y^{2}} & d_{xy}\\
d_{xy} & s-d_{x^{2}-y^{2}}\end{array}\right)$ & + & - & $A_{1}$\tabularnewline
\midrule 
\addlinespace[1mm]
2 & $\left(\begin{array}{cc}
s+d_{x^{2}-y^{2}} & g\\
g & -s+d_{x^{2}-y^{2}}\end{array}\right)$ & + & - & $B_{1}$\tabularnewline\addlinespace[1mm]
\midrule 
\addlinespace[1mm]
3 & $\left(\begin{array}{cc}
d_{xy}+g & s\\
s & d_{xy}-g\end{array}\right)$ & + & - & $B_{2}$\tabularnewline\addlinespace[1mm]
\midrule 
\addlinespace[1mm]
4 & $\left(\begin{array}{cc}
d_{xy}+g & d_{x^{2}-y^{2}}\\
d_{x^{2}-y^{2}} & -d_{xy}+g\end{array}\right)$ & + & - & $A_{2}$\tabularnewline\addlinespace[1mm]
\midrule 
\addlinespace[1mm]
5 & $\left(\begin{array}{cc}
0 & p_{x}\\
-p_{x} & 0\end{array}\right),\,\left(\begin{array}{cc}
0 & p_{y}\\
-p_{y} & 0\end{array}\right)$ & - & - & $E$\tabularnewline\addlinespace[1mm]
\midrule 
\addlinespace[1mm]
6 & $\left(\begin{array}{cc}
\alpha_{1}p_{x} & p_{y}\\
p_{y} & \alpha_{2}p_{x}\end{array}\right),\,\left(\begin{array}{cc}
\alpha_{2}p_{y} & p_{x}\\
p_{x} & \alpha_{1}p_{y}\end{array}\right)$ & + & + & $E$\tabularnewline\addlinespace[1mm]
\midrule 
\addlinespace[1mm]
7 & $\left(\begin{array}{cc}
0 & s\\
-s & 0\end{array}\right)$ & - & + & $A_{2}$\tabularnewline\addlinespace[1mm]
\midrule 
\addlinespace[1mm]
8 & $\left(\begin{array}{cc}
0 & d_{x^{2}-y^{2}}\\
-d_{x^{2}-y^{2}} & 0\end{array}\right)$ & - & + & $B_{2}$\tabularnewline\addlinespace[1mm]
\midrule 
\addlinespace[1mm]
9 & $\left(\begin{array}{cc}
0 & d_{xy}\\
-d_{xy} & 0\end{array}\right)$ & - & + & $B_{1}$\tabularnewline\addlinespace[1mm]
\midrule 
\addlinespace[1mm]
10 & $\left(\begin{array}{cc}
0 & g\\
-g & 0\end{array}\right)$ & - & + & $A_{1}$\tabularnewline\addlinespace[1mm]
\bottomrule 
\end{tabular}\caption{\label{tab:pairing}The possible pairing states. $P_{\mathrm{orbit}}$
($P_{\mathrm{spin}}$) denotes the parity of state in exchanging the
orbit (spin) index. I.R. shows the irreducible representation the
pairing state belongs to. Note that the table can also be constructed
by doing a symmetry classification of order parameter~\citep{Wang2008,Wan2008},
and the different pairing states belonging to the same irreducible
representation will in general mix. The schematic plots for some pairing
states are shown in Fig.~\ref{fig:pairing states}.}

\end{table}

\begin{figure}
\includegraphics[width=0.8\columnwidth]{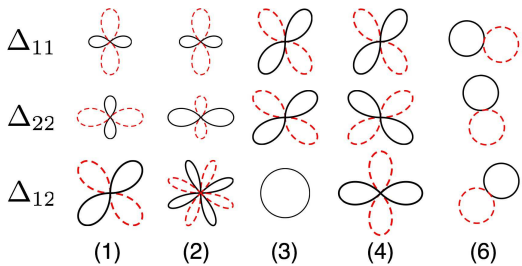}

\caption{\label{fig:pairing states}Schematic plot for multi-orbital-channel
pairing states ($P_{\mathrm{orbit}}=+1$). In plotting the states
(1) and (2), we have assumed that the $s$-wave component is smaller
than the $d$-wave components, due to the presence of the on-site
Coulomb repulsion.}

\end{figure}

In summary, we  have constructed the general effective interacting
Hamiltonian conforming to the peculiar symmetry of LaOFeAs system.
A complete list of the possible pairing states is also determined.
The general effective Hamiltonian can act as a maximal model, from
which the minimal model can be constructed for a given pairing state.
This is useful since the peculiar symmetry property of the electronic
state may render the first-sight intuitions misleading. The ten pairing
states put strong constraints to the possible forms of superconductivity,
and could act as the consistency check for the proposed pairing symmetries.
Finally, our approach is completely general. The result presented
here would be useful for other systems with the similar electronic
structure.

I thank Xi Dai for useful discussion on the band structure. This work
is supported by NSF of China No. 10734110, 10604063 and Ministry of
Science and Technology of China under 973 program No. 2006CB921304.

\bibliographystyle{apsrev}
\bibliography{/Users/shi/Documents/Papers/Papers}

\end{document}